\documentclass[a4paper,onecolumn,oneside,fleqn,10pt]{article}
\usepackage{abstract}
\usepackage{amsmath}
\usepackage{authblk}
\usepackage[english]{babel}
\usepackage{bm}
\usepackage[font={color=blue},figurename=Fig.,labelfont={it}]{caption}
\usepackage{cite}
\usepackage{enumitem}   
\usepackage{fchicago} 
\usepackage{float}
\usepackage[left=2cm,right=2cm,top=2cm,bottom=2cm]{geometry}
\usepackage{graphicx}
\usepackage{hyperref}
\usepackage{indentfirst}
\usepackage[utf8]{inputenc}
\usepackage[switch,columnwise]{lineno}
\usepackage{lipsum} 
\usepackage{pdfpages}
\usepackage{rotating}
\usepackage{setspace} 
\usepackage{subcaption}
\usepackage[varg]{txfonts}
\usepackage{wasysym}

\definecolor{blue}{RGB}{0,0,255}
\hypersetup{
    colorlinks=true,                          
    linkcolor=blue, 
    citecolor=blue, 
    urlcolor=blue,
    linktoc=all
}

\newcommand{\eg}{\textit{eg. }}
\newcommand{\cf}{\textit{cf. }}
\newcommand{\ie}{\textit{ie }}

\newcommand{\degree}{$^{\circ}$}

\usepackage{fancyhdr}
\graphicspath{{images/}{figures/}}

\usepackage{listings}
\usepackage{xcolor}  % Pour ajouter des couleurs

\begin{document}

\title{\huge On the Optimization of Singular Spectrum Analyses: A Pragmatic Approach \\}

% \Author[affil]{given_name}{surname}

\author[1]{Lopes Fernando}
\author[2]{Gibert Dominique}
\author[3]{Courtillot Vincent}
\author[3]{Le Mouël Jean-Louis}
\author[1]{Boulé Jean-Baptiste}

\affil[1]{\small UMR 7196 ,Muséum National d’Histoire Naturelle, Sorbonne Université, Paris, France}
\affil[2]{\small Académie des Sciences, Institut de France, Paris, France}
\affil[3]{\small UMR5276, Laboratoire de géologie de Lyon Terre, planètes et environnement (LGL-TPE), Lyon, France}

\date{\today}
\maketitle

\begin{abstract}
	Singular Spectrum Analysis (SSA) occupies a prominent place in the real signal analysis toolkit alongside Fourier and Wavelet analysis. In addition to the two aforementioned analyses, SSA allows the separation of patterns directly from the data space into the data space, with data that need not be strictly stationary, continuous, or even normally sampled. In most cases, SSA relies on a combination of Hankel or Toeplitz matrices and Singular Value Decomposition (SVD). Like Fourier and Wavelet analysis, SSA has its limitations. The main bottleneck of the method can be summarized in three points. The first is the diagonalization of the Hankel/Toeplitz matrix, which can become a major problem from a memory and/or computational point of view if the time series to be analyzed is very long or heavily sampled. The second point concerns the size of the analysis window, typically denoted as “L”, which will affect the detection of patterns in the time series as well as the dimensions of the Hankel/Toeplitz matrix. Finally, the third point concerns pattern reconstruction: how to easily identify in the eigenvector/eigenvalue space which patterns should be grouped. We propose to address each of these issues by describing a hopefully effective approach that we have been developing for over 10 years and that has yielded good results in our research work.	
	
	 \par\noindent\textbf{Keywords:} Singular Spectrum Analysis, Rank Reduction, Random SVD, Agglomerative hierarchical cluster tree
\end{abstract}

\section{\label{sec01} Introduction} 
	As is often the case, mathematical frameworks describing different types of signal analysis emerge only after their relevance to the problem at hand has been fully demonstrated. Medical imaging, like astronomical imaging, suffers from a loss of information caused primarily by two physical phenomena: diffraction and noise. The question is how to quantify this loss in order to reconstruct the diffraction image as accurately as possible. In many ways, diffraction can be represented by a Fredholm integral equation of the first kind, where the kernel represents the transfer function of the diffracting medium (such as the impulse response of an optical system). The resolving kernel, which characterizes the system's ability to resolve distinct features, plays a crucial role in understanding how diffraction limits the resolution of the reconstructed image (\eg \shortciteNP{menke1984,jourde2015}).  This is exactly what \shortciteN{bertero1982} found when they attempted to decompose this integral to a basis of singular functions, an approach theoretically founded by \shortciteN{miller1974}, who showed that such a decomposition was possible. Bertero and Pike's work uses singular value decomposition (SVD, Golub and Reinsch 1971) to understand how information loss limits the resolution of imaging systems by examining the energy distribution of the singular values. To address the ill-posed nature of the problem, the use of a generalized inverse allows for a stable reconstruction of the image by mitigating the effects of small singular values and reducing noise amplification (\eg \shortciteNP{penrose1955,hansen1998}). SVD is among the 10 most important algorithms of the 20th century (\cf \shortciteNP{cipra2000}). This process of analyzing the structure of the diffraction operator by studying its singular values allows for a better understanding of the resolution limits imposed by diffraction and noise. This is where the term “Singular” in Singular Spectrum Analysis (SSA) comes from. We illustrate this energy distribution in a different context, that of the Fourier transform, as discussed in \shortciteN{gibert2024}, Chapter 7, Section 1.2, pages 95-102.
	
	This method of decomposing information was later adopted by chaos specialists. To better understand the temporal evolution of the chaotic regime in the \shortciteN{lorenz1963} model, especially the transition to chaos, \shortciteN{broomhead1986} projected the one-dimensional time series into a two-dimensional delay matrix. This matrix consists of columns representing segments of the time series (delay vectors) and allows the observation of complex dynamics and the detection of hidden structures, such as strange attractors. Known as the trajectory matrix, this approach is used to reconstruct the trajectory in phase space of a dynamical system using temporal delays. Each column of the matrix represents a vector that is a time-shifted version of the original series. The key parameters in this process are the time delay  and the embedding dimension , which depend on the system under study. In fact, Broomhead and King's trajectory matrix is very similar to a Hankel matrix (for Hankel matrix \eg \shortciteNP{lemmerling2001}), where each element on a constant diagonal is identical, thus linking the columns or rows to time-shifted sequences of the original series. Although both matrices share this time-shifting structure, they serve different purposes. This is the origin of the terms “trajectory matrix” and “embedding step” that are commonly used in SSA-related work.
	
	Finally, this new analytical method is integrated into the field of geophysics, and more specifically into the domain of paleoclimatologists. The primary goal of \shortciteN{vautard1989} was not the extraction of qualitative dynamics from experimental flow data, as \shortciteN{broomhead1986} had done, but rather the decomposition of paleoclimatic time series to identify trends, cycles, and other underlying modes, which will be discussed further in this study. Thus, instead of reconstructing phase space, primarily using the method of nearest neighbors (geometric analysis), the geophysicists turned to spectral analysis by reconstructing the data space and extracting the principal modes present in the time series using the algebraic properties of Singular Value Decomposition (SVD).
	
	For more than a decade, we have tested and validated the SSA method in various geophysical contexts (\eg \shortciteNP{usoskin2016,lemouel2017,dumont2020,courtillot2022,lopes2022}). More recently, we have extended this approach to biophysical problems (\eg \shortciteNP{boule2024,lemouel2024}). In each case, the goal has been to extract the modes and pseudo-periods present in the analyzed time series while overcoming specific challenges associated with field data: high noise-to-signal ratios, discontinuities, or closely spaced periodicities that make separability difficult.
	
	To address these challenges, we have explored several advanced variants of SSA, such as Multichannel-SSA (MSSA, \eg \shortciteNP{oropeza2011}), Multitaper-SSA based on the Multitapers of \shortciteN{park1987}, and Circulant SSA (CSSA, \eg \shortciteNP{bogalo2021}). Although theoretically promising and effective in some practical cases, these approaches failed to overcome the inherent limitations imposed by the specific characteristics of the time series we studied, such as signal complexity and variability in experimental conditions. These variants, particularly CSSA, have recently gained traction in the biomedical field (\eg \shortciteNP{chaitanya2024,liu2024}), but in our view their direct application to real field data remains limited.
		
		To overcome the limitations and challenges encountered in recent years, we have developed a pragmatic methodological approach that integrates proven concepts from signal analysis into canonical SSA. This methodology, which we describe herein, is based on a functional and targeted adaptation of SSA, enriched with complementary tools. In Section \ref{secII.1}, we will recall the fundamental principles and key steps of SSA. In Section \ref{secII.2}, we will discuss the limitations of the standard method, followed by a presentation of our methodological solutions in Section \ref{secII.3}. Section \ref{secIII} will illustrate the application of our approach to various biophysical and geophysical datasets, highlighting its effectiveness across different contexts. Finally, in Section \ref{secIV}, we will conclude and propose future perspectives.

\section{The Pragmatic Approach}
	\subsection{A General Overview of SSA\label{secII.1}}
		We will not delve into the details of the method here and instead invite the reader to consult the reference work on the subject by \shortciteN{golyandina2018}.
		
		SSA can be summarized in four steps.
		
		The first and most critical step is to project the signal into a matrix, often of the Hankel or Toeplitz type (\cf matrices \ref{eq01}). This step is called the embedding step. Let $X = \{x_1, x_2, \ldots, x_N\}$ be a time series of length $N$ and let $L$ denote the size of the analysis window, the lag window mentioned in the introduction. In the case of a Hankel matrix, we obtain $K$ segments of length $L$, where $K=N-L+1$ 
\begin{equation} \label{eq01}
\begin{array}{c@{\hspace{2cm}}c}
\text{Hankel Matrix} & \text{Toeplitz Matrix} \\
\\
\begin{bmatrix}
x_1 & x_2 & x_3 & \cdots & x_K \\
x_2 & x_3 & x_4 & \cdots & x_{K+1} \\
x_3 & x_4 & x_5 & \cdots & x_{K+2} \\
\vdots & \vdots & \vdots & \ddots & \vdots \\
x_L & x_{L+1} & x_{L+2} & \cdots & x_N
\end{bmatrix}
&
\begin{bmatrix}
x_1 & x_2 & x_3 & \cdots & x_N \\
x_2 & x_1 & x_2 & \cdots & x_{N-1} \\
x_3 & x_2 & x_1 & \cdots & x_{N-2} \\
\vdots & \vdots & \vdots & \ddots & \vdots \\
x_N & x_{N-1} & x_{N-2} & \cdots & x_1
\end{bmatrix}
\end{array}
\end{equation}

	As shown in relations (\ref{eq01}), the Hankel matrix ($H$) has a structure characterized by constant descending diagonals (\ie $H[i,j] = x_{i+j-1}$), while the Toeplitz matrix ($T$) has a structure defined by repeated columns (\ie $T[i,j] = x_{abs(i-j)+1}$). An immediate implication is that the Hankel matrix is more general, while the Toeplitz matrix is better suited for detecting cycles or periodicities (\eg \shortciteNP{santhanam2000,hariprasad2024}). The advantage of these two matrices, especially the Hankel matrix, is that they preserve the temporal relationships within the time series. In the case of the Hankel matrix, the diagonals contain values that have similar temporal “lag”. This structure makes it possible to capture cycles, trends, and regularities in the series. As an example, in the case of the Fourier transform, which can be expressed as a matrix product, the associated Fourier matrix is of the Vandermonde type. In fact, the Fourier transform fits a least-squares model consisting of a sum of fixed-frequency sinusoids whose amplitude and phase are determined.. Unlike the Hankel matrix, which is constructed directly from the data values of the time series, the Fourier matrix is constructed from the complex exponentials that define the Fourier transform. This means that it is parameter-driven, relying on the frequencies, rather than data-driven like the Hankel matrix.
	
	The second step is the diagonalization of the matrix (Hankel or Toeplitz), for which the classical approach is to use the Singular Value Decomposition (SVD). We will briefly recall some of its principles. For a matrix \textbf{M} of dimensions $L \times K$, (not necessarily square), the SVD is given by,
\begin{equation}
	\textbf{M} = \textbf{U} \Sigma \textbf{V}^{t}
\end{equation}	

where,
\begin{itemize}
	\item \textbf{U} is an orthogonal matrix of size $L \times L$ ($\textbf{U}^{t}\textbf{U} = \mathcal{I}_{L}$) hose columns (the eigenvectors of $\textbf{MM}^{t}$) form an orthonormal basis of the row space of \textbf{M}. These vectors define the principal directions of variance in the original space.
	\item \textbf{V} is an orthogonal matrix of size $K \times K$ ($\textbf{V}^{t}\textbf{V} = \mathcal{I}_{K}$) whose columns (the eigenvectors of $\textbf{M}^{t}\textbf{M}$) form an orthonormal basis of the column space of \textbf{M}. These vectors define the principal directions in the target space. 
	\item $\Sigma$ is a diagonal matrix of size $L \times K$, whose diagonal elements ($\sigma_{1}, \sigma_{2}, \ldots$) are the singular values of \textbf{M}, ordered in descending order ($\sigma_{1} \geq \sigma_{2} \geq \ldots \geq 0$). These singular values measure the relative importance of the corresponding modes in the decomposition of \textbf{M}, quantifying the contribution of each singular mode to the structure of \textbf{M}. The singular values constitute the spectrum, which is the second 'S' in SSA.
\end{itemize}		
		
		Note that \textbf{M} is an operator mapping a vector space of dimension $K$ to a vector space of dimension $L$. Therefore, there are two sets of basis vectors, one for each vector space. However, only a subset of these vectors are related by the non-zero singular values, whose maximum number is $min(K,L)$. The remaining vectors are constructed via an orthogonalization process that does not use the information contained in the matrix and represents the null space of the operator.

			The third step, which is also critical, is to pair the eigentriplets (eigenvectors and singular values) obtained from the SVD that correspond to the same component of the original signal. In the case of time series, the eigenvectors associated with \textbf{U}, which represent the principal directions in the row space of the matrix \textbf{M}, are often interpreted as temporal modes. On the other hand, the eigenvectors associated with \textbf{V}, which represent the principal directions in the column space of the matrix, are generally interpreted as spatial or frequency modes.
			
			Here, we encounter the concept of orthogonal empirical functions, where the goal is to identify pairs of Hilbert transforms to account for the increasing phase shift between the columns of the Hankel matrix. By pairing these eigenvectors \textbf{U} and \textbf{V} with their corresponding singular values, one can selectively reconstruct certain parts of the signal using only the modes associated with the largest singular values. This approach allows for a more accurate reconstruction by compensating for phase shifts present in the signal components.
			
			This process enables the reconstruction of an accurate approximation of the original matrix and thus of specific components of the signal. The central argument for reconstructing specific signal components is that the SVD decomposes the matrix into modes that can be interpreted as independent components of the signal. The largest singular values correspond to the most significant directions, allowing the dominant component of the signal to be reconstructed. Conversely, the smaller singular values represent less significant contributions, often associated with residual components or noise. 
			
			The question of how to pair eigentriplets is a broad problem that often revolves around the phases of the components to be grouped and the Hilbert pairs. We will not cover this topic here, and instead refer the reader to \shortciteN{golyandina2001}, \shortciteN{hassani2007} and \shortciteN{bonizzi2014}.
			
			For example, if the first three eigentriplets correspond to the same component of the signal, then their combination by SVD allows reconstruction of an accurate approximation of that component and, by extension, an improved approximation of \textbf{M},
\begin{equation}
	\textbf{M} \approx \sigma_1 \textbf{U}_1 \textbf{V}_1^t + \sigma_2 \textbf{U}_2 \textbf{V}_2^t + \sigma_3 \textbf{U}_3 \textbf{V}_3^t
\end{equation}			

or in a more detailed form,
\begin{equation}
\begin{aligned}
\textbf{M} &\approx \sigma_1 \cdot 
\begin{bmatrix}
u_{11} \\
u_{21} \\
\vdots \\
u_{L1}
\end{bmatrix}
\cdot
\begin{bmatrix}
v_{11} & v_{12} & \cdots & v_{1K}
\end{bmatrix} \\
&\quad + \sigma_2 \cdot 
\begin{bmatrix}
u_{12} \\
u_{22} \\
\vdots \\
u_{L2}
\end{bmatrix}
\cdot
\begin{bmatrix}
v_{21} & v_{22} & \cdots & v_{2K}
\end{bmatrix} 
\quad + \sigma_3 \cdot 
\begin{bmatrix}
u_{13} \\
u_{23} \\
\vdots \\
u_{L3}
\end{bmatrix}
\cdot
\begin{bmatrix}
v_{31} & v_{32} & \cdots & v_{3K}
\end{bmatrix}.
\end{aligned}
\end{equation}		

	The final step is to reconstruct an approximation signal or the components obtained from the previous eigentriplets pairing. This step, known as “hankelization”, consists of computing the reconstructed matrix \textbf{M}, averaging over its diagonals (anti-diagonals in the SSA context, \cf relation \ref{eq03}), and using these values to reconstruct the corresponding time series ($x_t$),
\begin{equation}
	x_t = \frac{1}{|\{(i, j) : i + j = t + 1\}|} \sum_{i+j=t+1} M_{i,j},
\quad t = 1, \dots, N,
\label{eq03}
\end{equation}	
	
	\subsection{The Weaknesses of SSA \label{secII.2}}
	As we have just seen, the two most important points of the SSA method are the selection and construction of the trajectory matrix and the pairing of the eigentriplets.
	
	When $L$ is small, the Hankel matrix emphasizes the local details of the signal, favoring the capture of high-frequency components such as noise or small, fast oscillations. Conversely, long-term trends or longer periodic oscillations will not be well represented as they will exceed the scope of the small window. On the other hand, if $L$ is large, the window is better suited to represent slow components and global trends of the signal. Longer periodic oscillations can also be captured more effectively. In addition, signals with fractal behavior or long-term correlations (such as Hurst series) become more apparent with a large window, as correlations between distant segments are better represented. It should also be noted that the window length is crucial for accurately capturing non-stationarities in the signal. A window that is too short might fail to account for changes in the statistical properties of the signal over time, while an appropriately chosen window can effectively capture these variations. However, a large window may introduce more redundancy into the matrix as successive segments overlap, making it more difficult to separate different modes. All of these considerations explain the instability of SSA when it comes to decomposing a signal.
	
	For geophysical and biophysical considerations, the choice of a large window size ($L$) is often preferable, as it allows the capture of long-term trends and seasonality, while reducing the effects of noise inherent in field measurements. However, there are challenges associated with this approach. First, the amount of memory required to represent such a large matrix can become significant, especially for datasets sampled at high frequencies (eg. seconds) over several years (\cf \shortciteN{boule2024}). Second, the computational cost of performing SVD increases significantly with the size of the matrix, which may limit the feasibility of this approach for very large datasets. 	
	
	For an $L \times K$ ($L \geq K$), the computational complexity of the complete SVD algorithm is generally of the order of $\mathcal{O}(LK^{2})$ which can be decomposed as follows,
	\begin{itemize}[noitemsep, topsep=1pt, parsep=0pt]
		\item $\mathcal{O}(LK^{2})$ if $L \geq K$. This complexity dominates when decomposing a “wide” matrix.
		\item $\mathcal{O}(L^{2}K)$ if $L \leq K$. Dominates when decomposing a “tall” matrix.
	\end{itemize}

	These costs are due to the following steps,	
	\begin{itemize}[noitemsep, topsep=1pt, parsep=0pt]
		\item bidiagonal reduction: the matrix is first transformed into a bidiagonal form. This has a complexity of $\mathcal{O}(LK^{2})$ for "large" matrices.
		\item Jacobi or Golub-Kahan iterations: the final step of the SVD, which involves diagonalizing the bidiagonal matrix, has a complexity of $\mathcal{O}(K^3)$ in the worst case.
	\end{itemize}
	
	The complexity of an algorithm refers to the measure of its efficiency in terms of the resources required, such as computational time or memory space, required to solve a given problem as a function of the input size (here $N$).	

	There are alternatives to classical SVD to reduce these complexity, such as,
	\begin{itemize}[noitemsep, topsep=1pt, parsep=0pt]
		\item Truncated SVD: If only the $q$ largest singular values and their associated vectors are required ($q \ll min(L,K)$), the complexity can be reduced to $\mathcal{O}(Lq^{2}+q^{3})$ (\textit{eg} \shortciteNP{vannieuwenhoven2012,li2019})
		\item Randomized SVD: For large matrices, this method can be used to approximate the SVD with a reduced complexity of $\mathcal{O}(LK \ln q + q^{2})$, where $q$ is the desired approximate rank (\eg \shortciteNP{tamascelli2015,noorizadegan2024}).
		\item Sparse SVD: When the matrix is sparse (ie. contains many zero elements), the complexity depends on the number of nonzero elements, denoted as . In this case, the complexity can be reduced to $\mathcal{O}(nnz. q)$, where $q$ is the number of singular values extracted (\eg \shortciteNP{benjamin2017,grigori2018})
	\end{itemize}
	
	In the context of SSA, separability of components is essential to decompose a time series into distinct elements such as trend, cycles and noise. Weak separability, characterized by non-orthogonal singular sub-spaces and closely spaced eigenvalues, complicates this decomposition and hinders the interpretation of the results. \shortciteN{golyandina2013} proposed an alternative approach, based on non-orthogonal decompositions, to improve component discrimination under such conditions.
	
	Hilbert pairs, defined as principal oscillations and their quadrature components (phase-shifted by 90\degree), provide an analytical representation of oscillatory signals in terms of instantaneous amplitude and phase. In SSA, principal oscillations are often described by pairs of singular vectors analogous to Hilbert pairs. These pairs are necessarily eigentriplets (a set consisting of an eigenvalue and its associated singular vectors) whose eigenvalues are closely spaced. This proximity reflects their oscillatory nature, with an in-phase and a quadrature component. However, problems of weak separability can lead to misidentification of these pairs, compromising the accurate reconstruction of the phases and amplitudes of periodic cycles. These challenges highlight the importance of improving separation methods in the analysis of complex time series (\eg \shortciteNP{harmouche2017,hassani2017,vitali2019}).

	\subsection{Our Proposal for Improvement\label{secII.3}}
	The corresponding MATLAB code can be found in Appendix \ref{app:A}.

	\subsubsection{Signal extension\label{sec03a}} 
	As previously discussed, one of the two main limiting factors of the SSA algorithm is the complexity of the SVD of a trajectory matrix. For our applications, it is desirable that the dimension L is as large as possible. We choose to work with a Hankel matrix which, as mentioned earlier, is sufficiently general for our purposes. However, as our matrices (\ref{eq01}) show, the segments of the time series located at the boundaries of \textbf{H} (\eg $\begin{bmatrix} x_1 \\ x_2 \\ x_3 \\ \vdots \\ x_L \\ \end{bmatrix} $ and $\begin{bmatrix} x_{K} \\ x_{K+1} \\ x_{K+2} \\ \vdots \\ x_N \\ \end{bmatrix}$) are less represented during the embedding process than the other segments. This naturally leads to difficulties in detecting and correctly extracting features, a phenomenon that can be described as a boundary effect. A classic method in geophysics, inspired by the zero padding approach, is to duplicate the initial and final segments of the signal at its beginning and end. By increasing the dimension of  \textbf{H} in this way, the boundary segments of the original time series are shifted to positions where they are more prominent in \textbf{H}, in the hope that they will be better represented.

\begin{figure}[ht!]
    \centering
    \includegraphics[width=\textwidth]{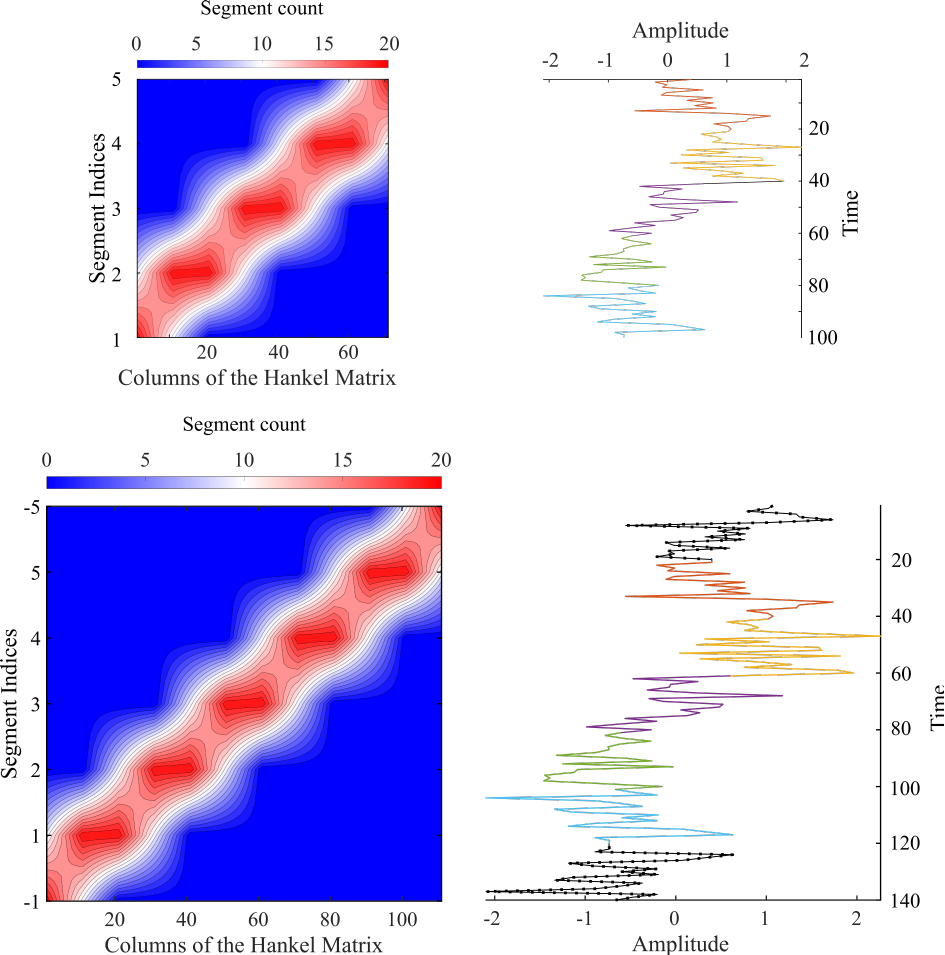}  
    \caption{On the right is the frequency at which a colored segment of a time signal (right column) appears in the Hankel matrix. This is shown at the top for the original signal and at the bottom after the signal has been extended by duplicating its boundary segments.}
    \label{fig:01}
\end{figure}

	In Figure \ref{fig:01}, we illustrate this data extension process and its effect on the Hankel matrix. At the top right of Figure \ref{fig:01}, a time signal ($N=100$) is shown, consisting of a noisy sinusoidal waveform. This signal is divided into segments of size $P=20$, resulting in five colored segments. Below, the same signal is extended by duplicating and symmetrizing the boundary segments ($i=1$ and $i=5$), which are mirrored to avoid discontinuities. These additional segments are represented as dashed black lines in the bottom right of Figure \ref{fig:01}. On the left, the corresponding matrices are displayed. The column indices ($K$) of the Hankel matrix are represented on the x-axis. For generalization, we selected  $L=30$, which reflects a practical scenario where the existence of five segments in the signal is known, but their exact dimensions are not. For the original signal (top matrix), $K$ is calculated as $K=100-30+1=71$. For the extended signal (bottom matrix), it becomes $K=100+2*20-30+1=111$.
	
	As observed, each segment is could be counted up to 20 times per column, and the informational support of each is visualized as red ellipsoids in the matrix. In the top matrix, three out of five ellipsoids are complete, indicating potential boundary effects. In the bottom matrix, corresponding to the extended signal, all five ellipsoids are correctly represented, though the duplicated segments appear incomplete.
	
	\subsubsection{Randomized-SVD}	
	This first step, which increases the presence of the end segments, also increases the size of the Hankel matrix, which can be significant in our applications. To diagonalize this matrix, we have chosen to use Randomized-SVD, a reduced-rank approximation method that serves as an efficient alternative to exact SVD (\eg \shortciteNP{erichson2016,darnell2017,alla2019,nakatsukasa2024}). Randomized-SVD builds upon the work of \shortciteN{frieze2004}, which lays the foundation for random-based matrix factorization methods, and introduces probabilistic approaches for low-rank approximations. This approach has several advantages. First, it significantly reduces the computational time, making it faster than exact SVD. Second, it is particularly suitable for large matrices due to the preliminary dimension reduction, which provides excellent scalability. Finally, it ensures robust approximations of the dominant singular values while maintaining high accuracy in the final result. The latter correspond to trends and pseudo-cyclicities that may be of interest in geophysics and biophysics.
	
	Regarding the complexity of the Randomized-SVD, let us recall that the method consists of projecting the matrix \textbf{H} onto a lower-dimensional subspace $q$; where $q$ is the desired approximate rank. For a matrix of size $L \times K$ and a reduced rank $q$, such that $q \ll min(L,K)$, the complexity of the method is $\mathcal{O}(LKq) + \mathcal{O}(q^{2}(L+K))$. Here,$\mathcal{O}(LKq)$ corresponds to the phase of projecting the matrix onto a subspace of dimension $q$, and $\mathcal{O}(q^{2}(L+K))$ corresponds to the computation of the exact SVD on the smaller matrix obtained after projection. Here is a schematic representation and pseudo-code corresponding to a Randomized-SVD,
\begin{enumerate}
	\item construction of the Hankel matrix (\textbf{H}) of size $L \times K$;
	\item reduced rank definition: $rank = min(q,L,K)$;
	\item initialization of \textbf{Q} matrix (size $L \times q$) with values drawn from a standard normal distribution $ \mathcal{N}(0,1)$; $\textbf{Q} = randn(L,q)$. This matrix will serve as a starting point to capture the principal directions of \textbf{H};
	\item a power iterations loop performs several iterations of the power method to improve the approximation of the dominant sub-spaces of \textbf{H}. At each iteration, \textbf{Q} is multiplied by \textbf{H} and by the transpose of \textbf{H}.$\textbf{Q} = \textbf{H}*(\textbf{H}^{'}*\textbf{Q})$ (\eg \shortciteNP{noorizadegan2024}).  This step amplifies the contributions of the dominant singular values;
	\item a $QR$ factorization of \textbf{Q} orthonormalizes the columns of \textbf{Q}. The decomposition is economic (compact), which means that only the necessary part of \textbf{Q} is retained;
	\item a projection of \textbf{H} on reduced space onto the column of \textbf{Q}, resulting in a matrix \textbf{B} of size $q \times K$. This step reduces the dimensions of \textbf{H} while preserving the important information $\textbf{B} = \textbf{Q}^{'}*\textbf{H}$
		\item SVD decomposition of \textbf{B} and reconstruction of singular vectors of\textbf{ H}.
\end{enumerate}	

	Let us briefly review the $QR$-decomposition (\cf \shortciteNP{francis1961,francis1962}). For a matrix \textbf{A} of size $m \times n$ where $m\geq n$, the $QR$-decomposition expresses \textbf{A} as \textbf{A} = \textbf{QR}. \textbf{Q} is an orthogonal matrix of size $m \times n$ (\ie $\textbf{Q}^{'}\textbf{Q} = \mathcal{I}$), with columns that are orthonormal. \textbf{R} is an upper triangular matrix of size $n \times n$. The complexity of such a decomposition is $\mathcal{O}(mn^{2})$. As we have seen, $QR$-decomposition plays a crucial role in Randomized-SVD because: i) After performing a random projection of the original matrix \textbf{H} using a random matrix \textbf{Q}, it is necessary to orthonormalize the columns of \textbf{Q} and ii) the $QR$-decomposition reduces the dimensionality of the problem to $q \times K$. Moreover, it serves an even more important function: numerical stabilization. The orthonormalization provided by the $QR$-decomposition numerically stabilizes the algorithm by preventing collinearity issues among the vectors in the matrix \textbf{Q}. This ensure that the approximations of the dominant singular vectors are reliable (\eg \shortciteNP{martinsson2019,nakatsukasa2020}).  
	
\begin{figure}[ht!]
    \centering
    \includegraphics[width=\textwidth]{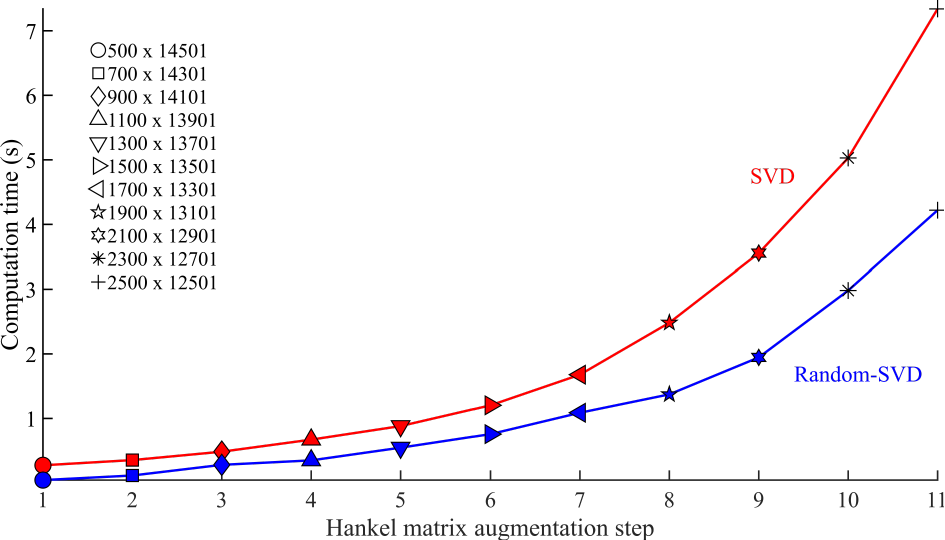}  
    \caption{Evolution of computation time (y-axis) as a function of the increasing complexity of the Hankel matrix (x-axis). The chosen rank ($q$) is arbitrarily set to  $L=300$ points. In red, the computation time for the canonical SVD, and in blue, the computation time for the Randomized-SVD with two iterations of the power loop.}
    \label{fig:02}
\end{figure}

	Figure \ref{fig:02} illustrates these observations. Using a synthetic signal (of size $N = 15,000$ points), we constructed and diagonalized the corresponding Hankel matrix for window sizes  ranging from 500 to 2500. We deliberately chose small sizes for pedagogical purposes to keep the computation time below ten seconds. For the Randomized-SVD, we arbitrarily reduced the rank to  and chose two iterations for the power loop. The computations were performed on a standard laptop (16 x Intel(R) Core(TM) i7-10870H CPU @ 2.20GHz – 32GB RAM) using MATLAB 2024b. We used the software's native functions: \textit{hankel()}, \textit{svd()}, \textit{randn()}, and \textit{qr()}.
	
	\subsubsection{Eigenvalues thresholding\label{sec03c}}	
\begin{figure}[ht!]
    \centering
    \includegraphics[width=\textwidth]{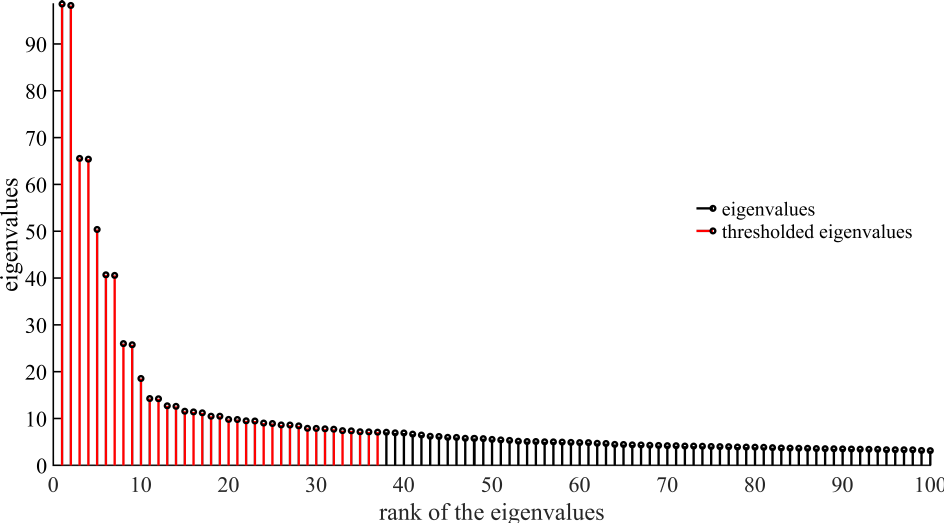}  
    \caption{Singular values obtained after the SVD of the Hankel matrix of the polar motion time series (in black). In red, the singular values whose the cumulative sum exceeds the threshold, set to 90\%.}
    \label{fig:03}
\end{figure}

	As mentioned earlier, the SVD algorithm orders these eigenvectors in descending order of their respective singular values. Thus, it is possible to draw an analogy between the sum of the singular values and the energy corresponding to the sum of the components of a given signal. The thresholding we apply is quite simple: we retain all singular values (and their corresponding singular vectors) whose cumulative sum exceeds a predefined threshold. As an illustration, Figure \ref{fig:03} shows the first 100 singular values (\cf $\Sigma$ ) obtained after diagonalizing the matrix \textbf{H}, constructed from the first component of the time series of the rotational polar motion. We will revisit this time series in the next section. In black are the singular values before thresholding, and in red are the singular values selected after applying a 90\% threshold. In other words, 90\% of the energy of this time series is represented by the 37 red singular values (and their corresponding singular vectors).
	
	\subsubsection{Reconstruct individual eigenvectors associated with thresholded eigenvalues \label{sec:03d}}
	At this stage, the signals components corresponding to a single, and only a single eigentriplet ($U_{i},\Sigma_{i}, V_{i}$) are reconstructed for all eigentriplets retained after the thresholding in the previous section \ref{sec03c}. The signal extensions from part \ref{sec03a} are removed.
	
	\subsubsection{Clustering the reconstructed components}
	The set of thresholded eigentriplets are individually reconstructed so that the grouping of the signal components occurs in the data space rather than in the dual space. The approach follows the principle of Hilbert pairs.
	
	First, the correlation matrix ($\rho$) between the different signal components reconstructed in section \ref{sec:03d} is computed. Then, $\rho$ is transformed into an average distance matrix ($1-abs(\rho)$) to group components with high correlations. Finally, hierarchical clustering is applied using the average linkage method.
	
	Hierarchical clustering with average linkage works in an agglomerative manner, starting with each object as an individual cluster and progressively merging clusters based on their similarity until a single cluster is formed (\eg \shortciteNP{murtagh2012,saraccli2013,ros2019,jarman2020}). In the case of average linkage, the distance between two clusters is defined as the average of the distances between all points in the two clusters. More formally, for two clusters $A$ and $B$, the distance $d$ is given by,
\begin{equation}
		d(A,B) = \dfrac{1}{|A|.|B|} \sum_{i\in A}\sum_{j\in B} d(i,j)
		\label{eq04}
\end{equation}	 	
	
	where $|A|$ and $|B|$ are the number of elements in clusters $A$ and $B$, respectively, and $d(i,j)$ is the distance between the points $i$ and $j$. Each object (\ie each signal component) is initially considered as a single cluster. In the first step, the distance between all clusters is computed using the average linkage (the average of the distances between points in the clusters). In the second step, the two clusters with the smallest average distance are merged, and the distances between the newly merged cluster and the remaining clusters are recalculated. This process of merging and updating clusters is repeated until only one cluster remains. The result can be displayed as a dendrogram, and the components within the same cluster can be reconstructed by summing them. In MATLAB, all of this can be summarized by the functions \textit{linkage()} and \textit{cluster()}.
	
	We will now use real data to illustrate the results of our pragmatic approach, first on the time series of the polar motion, and then on continuous temperature measurements of trees in Paris.
	
\section{Two Examples: one in Geophysics and one in Biophysics\label{secIII}}
	\subsection{On the polar motion\label{sec:polar}}
\begin{figure}[ht!]
    \centering
    \includegraphics[width=\textwidth]{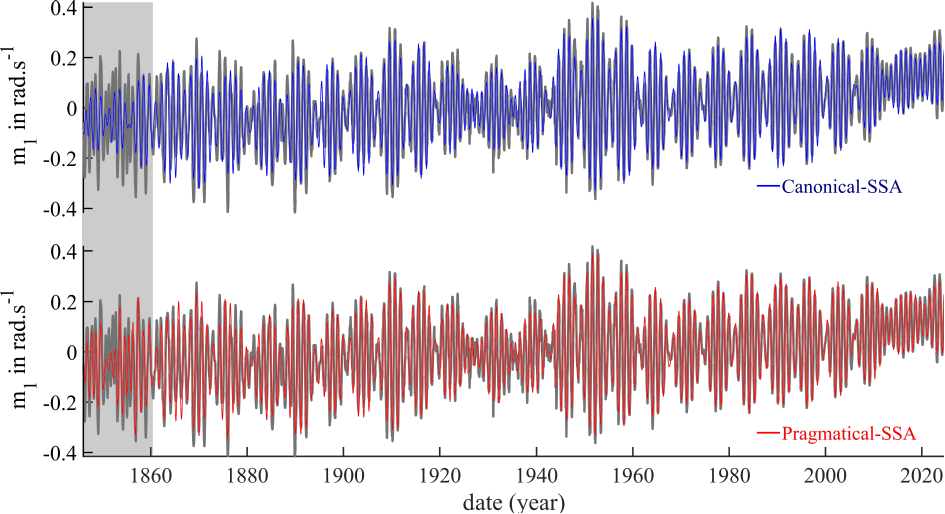}  
    \caption{The $m_1$ component of the polar motion (gray curve). At the top, overlaid in blue, is the component reconstructed using all the eigentriplets obtained after a canonical SSA. At the bottom, overlaid in red, is the component reconstructed using all the eigentriplets obtained after a pragmatic SSA. The gray shaded region from 1846 to 1860 represents a recently added segment to the $m_1$ time series. As can be seen, the reconstruction in this region is less accurate for both approaches.}
    \label{fig:04a}
\end{figure}

	Polar motion refers to the shift of the Earth's axis of rotation relative to the Earth's surface. The parameters $m_1$ and $m_2$ are used to quantify this shift. They represent the coordinates of the instantaneous pole (ie the Earth's axis of rotation at a given moment) expressed in a fixed reference frame tied to the Earth's crust. m1 represents the shift of the pole in the direction of the Greenwich meridian (the X axis in the Earth's reference frame). $m_2$ represents the shift of the pole in the direction of 90° relative to the Greenwich meridian (the Y axis in the Earth's reference frame). The time series (EOP C01 IAU1980) is provided by the IERS1. Only the $m_1$ (\cf Figure \ref{fig:04a}) component is considered here.
	
\begin{figure}[ht!]
    \centering
    \includegraphics[width=\textwidth]{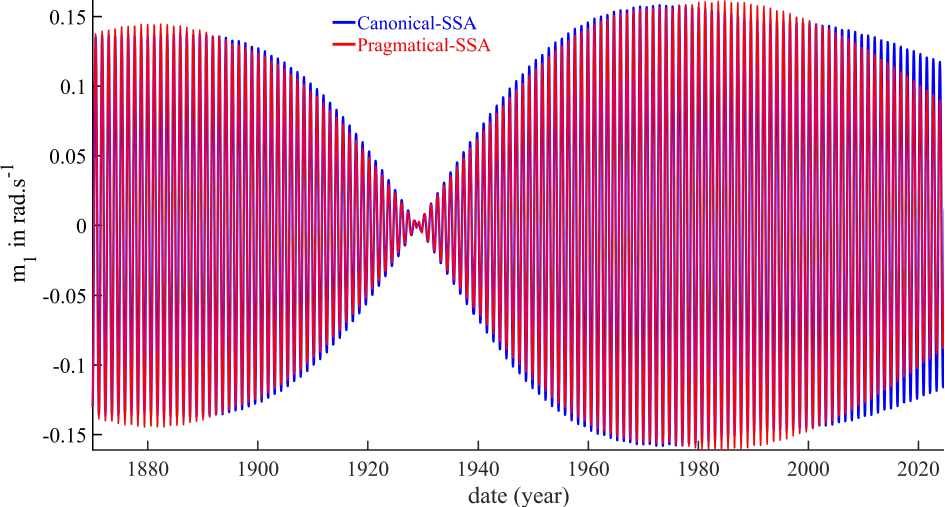}  
    \caption{The Chandler wobble extracted from the data shown in Figure \ref{fig:04a}; in blue is the canonical-SSA, in red is our pragmatic-SSA.}
    \label{fig:04b}
\end{figure}

\begin{figure}[ht!]
    \centering
    \includegraphics[width=\textwidth]{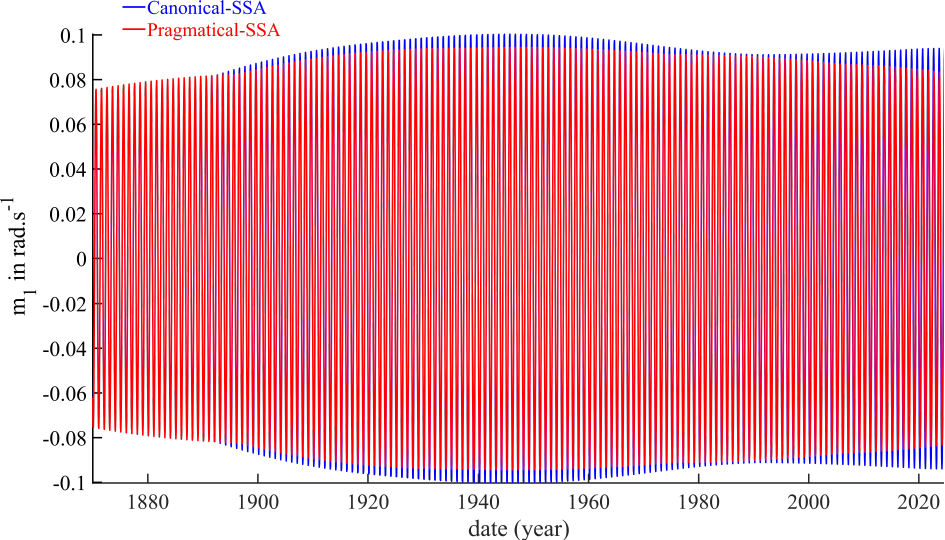}  
    \caption{The seasonal oscillation extracted from the data shown in Figure \ref{fig:04a}; in blue is the canonical-SSA, in red is our pragmatic-SSA.}
    \label{fig:04c}
\end{figure}

\begin{figure}[ht!]
    \centering
    \includegraphics[width=\textwidth]{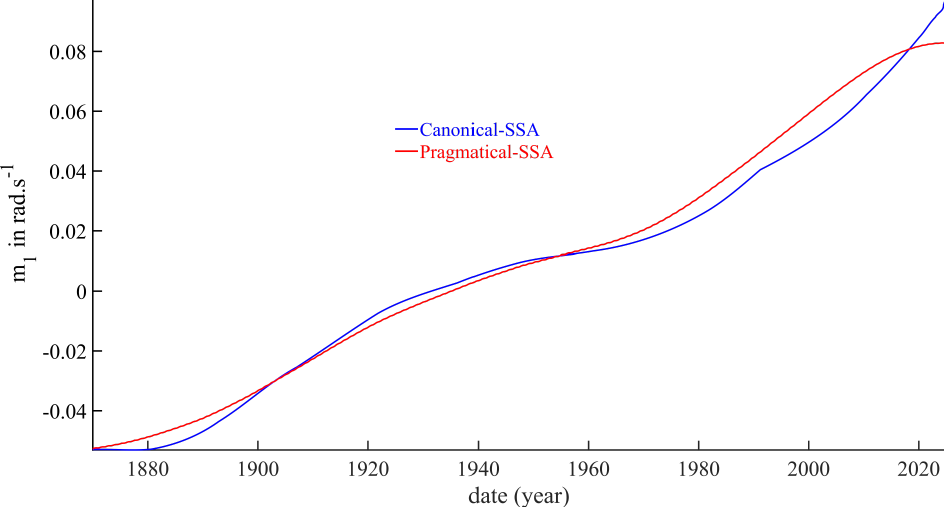}  
    \caption{The polar drift extracted from the data shown in Figure \ref{fig:04a}; in blue is the canonical-SSA, in red is our pragmatic-SSA.}
    \label{fig:04d}
\end{figure}

	To first order, the $m_1$-component can be decomposed (\eg \shortciteNP{lopes2021}), in order of importance, into the sum of the Chandler free oscillation (\cf \shortciteNP{chandler1891a,chandler1891b}, \cf Figure \ref{fig:04b}), the forced annual oscillation (\cf \shortciteNP{lambeck2005}, \cf Figure \ref{fig:04c}), and the trend (\cf \shortciteNP{markowitz1968,stoyko1968}, \cf Figure \ref{fig:04d}).

	Figure \ref{fig:04a} shows the $m_1$-component of the polar motion since 1846. This component was analyzed using both a canonical-SSA and our pragmatic-SSA approach. From the sum of all computed eigentriplets, two approximations of $m_1$ were reconstructed: the upper curve in red represents the canonical-SSA, while the lower curve in blue corresponds to the pragmatic-SSA. A gray-shaded region highlights the time interval where both approaches fail to accurately reconstruct the original signal. This discrepancy occurs because this segment was recently added to the dataset by the IERS. The signal consists of 3577 data points.
	
	For the canonical-SSA, we chose a window size of $L=2900$, resulting in a Hankel matrix of dimensions $2900 \times 678$. Under these conditions, the three components mentioned above are captured by eigentriplets $i=\{1,2\}$ for the Chandler oscillation, $i=\{3,4\}$ for the annual oscillation, and $i=\{5\}$ for the Markowitz-Stoyko trend. Additional pseudo-periodic components are present, but we chose to limit the reconstruction to the first 10 eigentriplets out of the 678 available to produce the blue curve in the top panel of Figure \ref{fig:04a}.
	
	For the pragmatic-SSA, we chose a window size $L$=1100 and a rank $q$=1000, retaining only the eigentriplets whose cumulative sum accounts for more than 90\% of the signal energy. The algorithm identified 38 eigentriplets, grouped into 23 clusters. The first three clusters correspond, in order, to the Chandler oscillation, the annual oscillation, and the trend. We used all 23 clusters to reconstruct the red curve shown in the bottom panel of Figure \ref{fig:04a}.  
	
		The three main components of polar motion extracted by the two SSA approaches are shown in Figure \ref{fig:04b} for the Chandler oscillation, Figure \ref{fig:04c} for the forced oscillation, and Figure \ref{fig:04d} for the drift of the rotational pole.

	The two approaches yield essentially identical results, and there is no fundamental criticism of either. This is a crucial point because, unlike Fourier or wavelet analysis, there is no fixed projection basis. Instead, the basis changes according to the structure of the Hankel matrix. In this sense, SSA benefits from an ad hoc projection tailored to the data.
	
		The only observations we can make are that in the three figures the behavior of the extracted components is largely consistent, their amplitudes are comparable, and the phases are preserved. It is noteworthy that among these three figures, only the Chandler cycle shows significant differences after the year 2000. It is important to note that the Chandler oscillation consists of two spectral peaks: one at 1.17 years and the other at 1.19 years. Starting in 2000, the phase of the blue curve obtained by the canonical-SSA drifts to 1.20 years. This drift is effectively captured by the eigentriplets $i=\{1,2\}$ In contrast, our pragmatic-SSA approach only captures the 1.17 and 1.19 year cycles. Whether this is correct or not is open to interpretation. Since the study of this free oscillation remains an active topic of discussion, we will refrain from further comments. 
		
\begin{figure}[H]
    \centering
    \includegraphics[width=\textwidth]{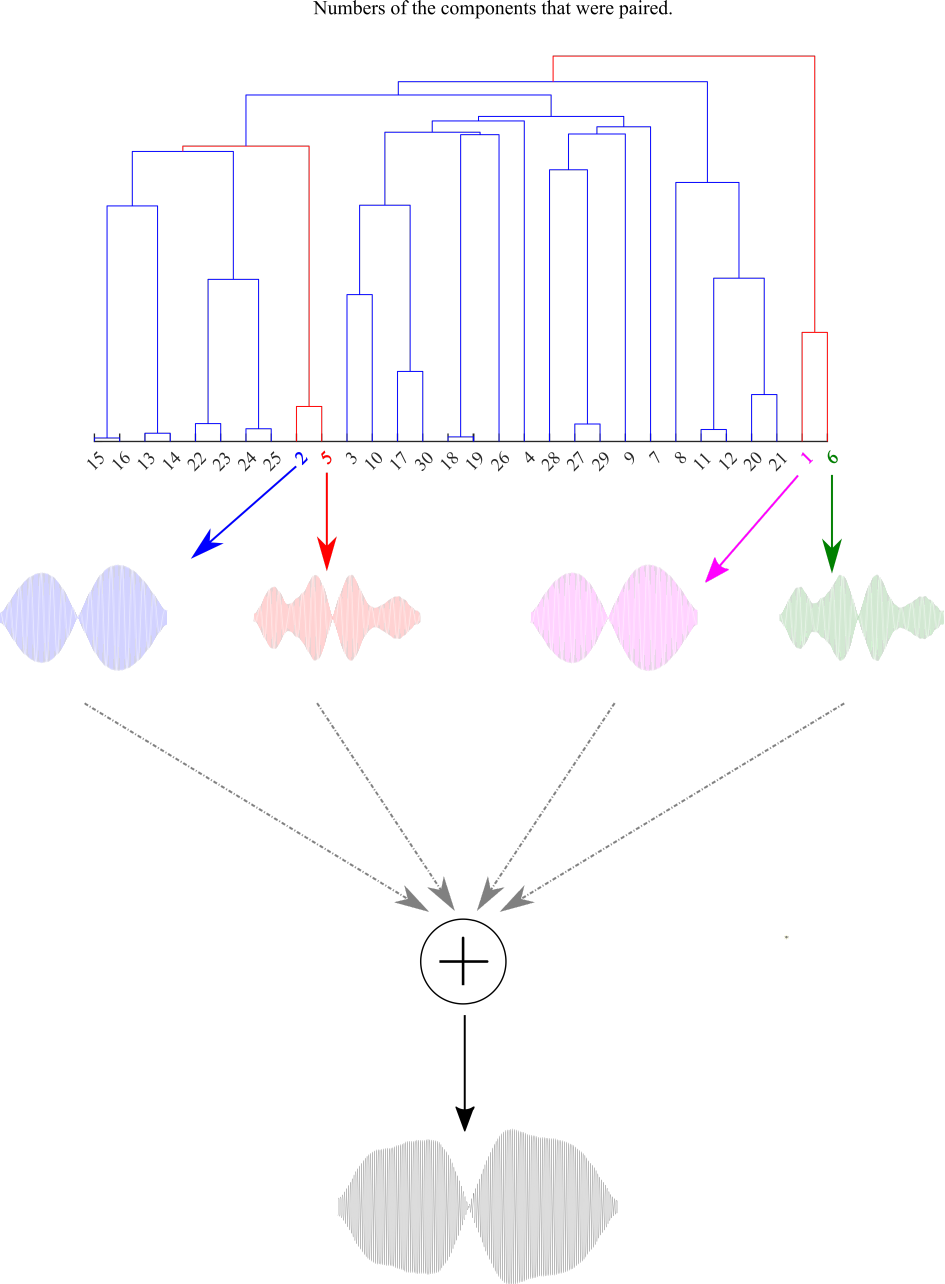}  
    \caption{At the top, the dendrogram of the pragmatic-SSA, which led to the results shown in Figures \ref{fig:04a}, \ref{fig:04b}, \ref{fig:04c}, and \ref{fig:04d}. The eigentriplets paired during the first step, as well as their respective components in the data space, have been colorized. At the bottom, in black, the sum of these components at step 1, already providing a good approximation of the Chandler oscillation.}
    \label{fig:04e}
\end{figure}

		Hierarchical clustering, through the construction of a dendrogram, allows us to track which components were grouped together and how they were merged during the clustering process. In Figure \ref{fig:04e} we present this dendrogram and illustrate how, in the first steps, the individual components corresponding to the eigentriplets $i=\{2,5\}$ and $i=\{1,6\}$ were paired. We display the corresponding components in the data space, using the same color as the eigentriplets numbers, and then sum them up. It can be observed that at the stage where we stopped for this plot, a good approximation of the Chandler oscillation is already reconstructed (black curve).

	\subsection{On trees in a Parisian grove\label{sec:trees}}
	Following previous research on tree bio-electricity (\cf \shortciteNP{gibert2006}), which we have recently revisited (\cf \shortciteNP{lemouel2024}), we will begin instrumenting a group of trees in a Parisian grove (France) in 2019 (\cf \shortciteNP{boule2024}). The study will focus on about ten trees, including hornbeams and oaks. Each tree will be equipped with four electrodes and four thermocouples (Pt-100) that will record bio-electric and bio-thermal data every second. In addition, reference electrodes and thermocouples will be installed in the root systems of these trees. This setup will generate approximately $\pi \times 10^{7}$ data points per sensor per year.
	
	Since the goal is to identify electrokinetic phenomena caused by tidal potential variations as well as potential communication signals, it is necessary to analyze the signals in their entirety using the largest possible window size. One of the challenges discussed in \shortciteN{lemouel2024} is that several tidal potentials are closely spaced. For example, among the diurnal tides, we have the \textbf{O1} tide (main lunar) with a period of 25 hours 49 minutes, the \textbf{K1} tide (lunar-solar declination) with a period of 23 hours 56 minutes, the \textbf{P1} tide (main solar) with a period of 24 hours 04 minutes, and the \textbf{Q1} tide (main lunar elliptical) with a period of 26 hours 52 minutes. Therefore, we must accurately separate these luni-solar tidal components.
	
\begin{figure}[ht!]
    \centering
    \includegraphics[width=\textwidth]{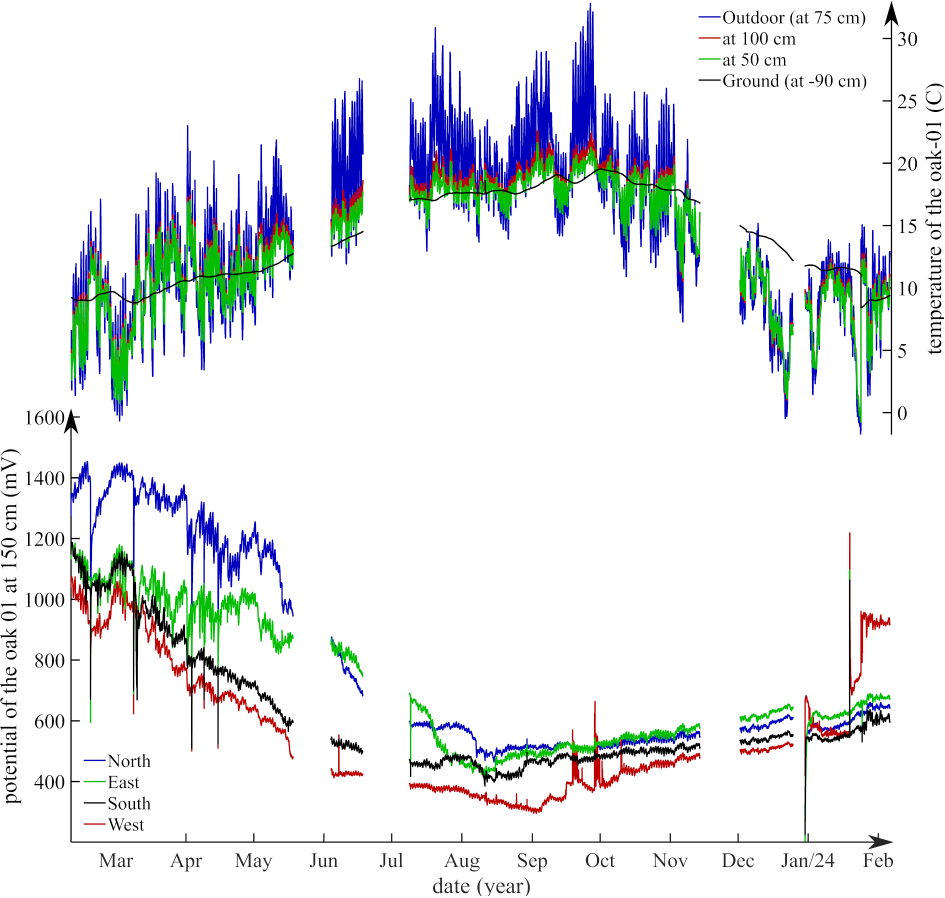}  
    \caption{From \protect\shortciteN{boule2024}. Temporal evolution since 2023: at the top, temperatures recorded within oak 01 at 50 cm and 100 cm (red curve), as well as in the soil at 90 cm depth (black curve), and outside the tree (attached to the tree) at 75 cm (blue curve); at the bottom, electrical potentials measured at 150 cm on the North direction in blue, on the East direction in green, on the South in black, and on the West in red.}
    \label{fig:05a}
\end{figure}

	\shortciteN{boule2024} discuss the relationship between electrokinetic phenomena induced by sap flow and the thermal regulation of six trees (three hornbeams and three oaks). In fact, trees can be considered as approximately vertical cylinders composed of thousands of channels (the xylem) and can thus be interpreted as heat exchangers. Figure \ref{fig:05a}, which corresponds to Figure 03 in \shortciteN{boule2024}, illustrates this mirrored relationship between electrical potential and thermal regulation in an oak over the course of a year.	
	
	In addition to this mirrored behaviour of the trends in the two biophysical signals (electrical and thermal), the most important primary forcing is the diurnal variation. However, as we have seen, there are several closely related lunisolar tides around the 24-hour period. It is important to preserve the phases and avoid spreading information from one tide to another. In \shortciteN{boule2024} we showed, using canonical-SSA, that the daily thermal oscillation is mainly driven by the \textbf{S1} tide (24h) from January to May, by the \textbf{K1} tide (23h56) in June and July, by the \textbf{S1} tide again from July to October, and by the \textbf{O1} tide (25h49) from November to December.
	
	Of course, the separation of these pseudo-cycles is mainly due to the temporal sampling rate of 1 second. At this stage, one question remains regarding our pragmatic-SSA approach: does the addition of mirrored segments to the original signal introduce phase shifts during reconstruction? Intuitively, it seems reasonable to assume that it does not, and that this process merely changes the order of the Hilbert pairs for a given oscillation. However, it is crucial to verify this assumption.
	
\begin{figure}[ht!]
    \centering
    \includegraphics[width=\textwidth]{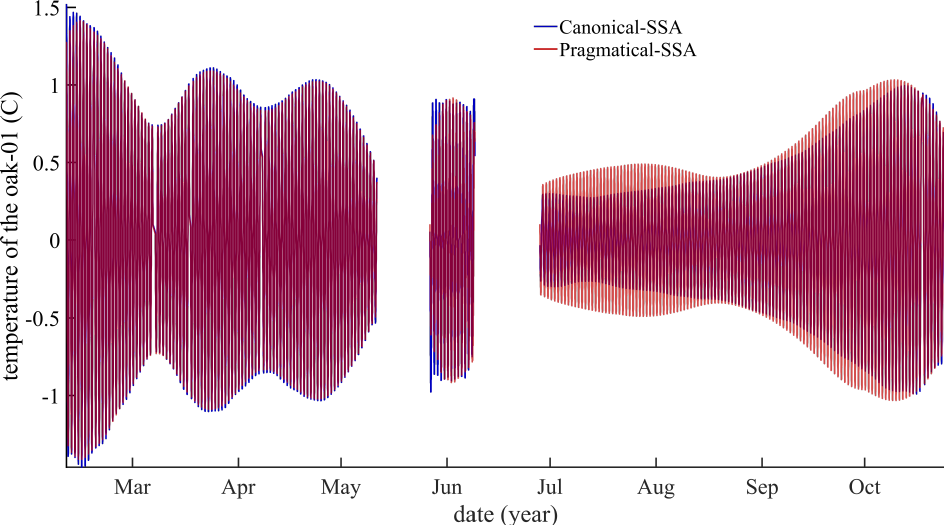}  
    \caption{Different diurnal components (\textbf{S1}, \textbf{O1}, and \textbf{K1}) extracted during the year 2023 using canonical-SSA (blue curve) and pragmatic-SSA (red curve).}
    \label{fig:05b}
\end{figure}

	In \shortciteN{boule2024}, we had to downsample the raw signals in order to apply our canonical-SSA. In the example presented here, we did not downsample the data. As in the paper, we performed SSA on segmented portions of the data due to important measurement interruptions, as shown in Figure \ref{fig:05a}. Each SSA was performed on (almost) continuous segments of the signal. For the extraction of the diurnal oscillation, we compare the results of the two analyses performed with the same window size ($L$) of 10 hours. The results are presented in Figure \ref{fig:05b} and can be compared with the identical results presented in Figure 05 of \shortciteN{boule2024}. Although it should be noted that pragmatic-SSA is essentially based on truncated SVD, the results are relatively satisfactory for a comparable computational time. Both approaches extract approximately the same information from February 2023 to May 2023. In June 2023, edge effects appear in the canonical-SSA (blue curve). For the last segment, from July to November 2023, the envelope obtained with our approach (red curve) shows slightly more modulation compared to the canonical-SSA (blue curve).
	
\begin{figure}[ht!]
    \centering
    \includegraphics[width=\textwidth]{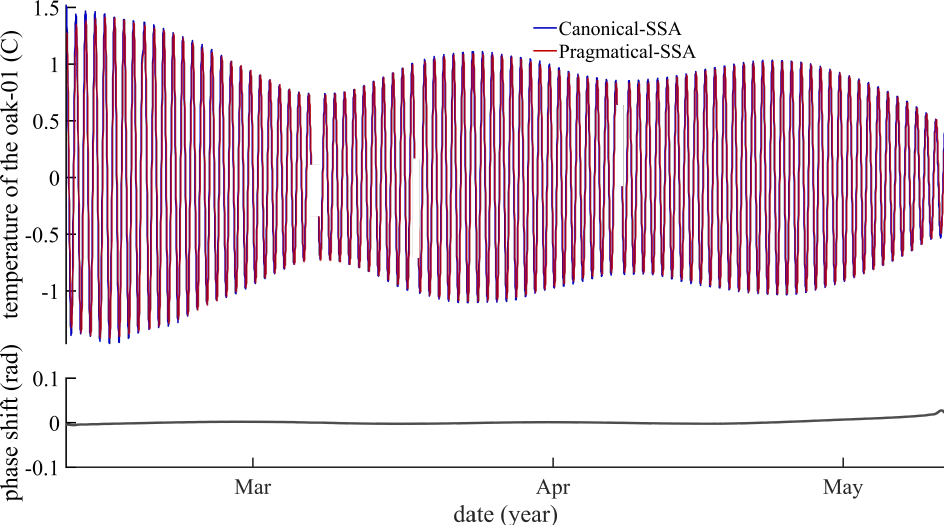}  
    \caption{Zoom on the results shown in Figure \ref{fig:05b}. The top panel shows the almost perfect superposition of the diurnal oscillation obtained by the two methods. The lower panel shows the instantaneous phase shift (in radians) between the extracted components.}
    \label{fig:05c}
\end{figure}

	Figure \ref{fig:05c} shows an enlargement of the results from Figure \ref{fig:05b}, covering the period from February to June 2023. It is clear that the curves overlap almost perfectly in both phase and amplitude. In the lower panel we have calculated the instantaneous phase shift (in radians) between the two pseudo-cycles obtained by the two approaches. As can be seen, this phase shift is close to zero over the whole period. This confirms the accurate phase reconstruction obtained by the pragmatic-SSA.

\section{Discussion and Conclusion\label{secIV}}
		Singular Spectrum Analysis (SSA) is undoubtedly one of the key tools in signal analysis, alongside Fourier and Wavelet analysis. It is characterized by its ability to detect and separate pseudo-cycles and non-stationary signals in real data. Its main drawback is that this capability relies on the properties of Hankel/Toeplitz matrices constructed (\textit{ad hoc}) from the data and on the SVD algorithm. While SVD is very effective, it is also very demanding in terms of memory and computational time.
		
		In the first part (\cf Section \ref{sec01}), we reviewed the historical context that led to the development of this new tool, starting with the physical-optical experiments of \shortciteN{bertero1982} and ending with the geophysical challenges, including the palaeoclimate studies of \shortciteN{vautard1989}. In the second part, we introduced and described the basic concepts of the canonical method (\cf Setion \ref{secII.1}) as well as its limitations (\cf Section \ref{secII.2}). These limitations can be trivially summarized as follows: how can increasingly long and finely sampled time series be efficiently analyzed within a reasonable time frame and using a standard computer system? As we have seen, there are numerous articles in the literature dealing with this question, but they all aim at finding an alternative to the Hankel matrix.
		
		We have pragmatically chosen to keep the matrix structure that seems most appropriate for SSA and to focus as much as possible on optimizing its diagonalization. Our approach, which we have called pragmatic-SSA (\cf Section \ref{secII.3}), is based on the following well-established concepts, 
\begin{itemize}
	\item  Redundancy by symmetrizing the edge segments of the signal to improve their representation within the Hankel matrix.
	\item  Reducing computational time and problem complexity through randomized-SVD. This method approximates the Hankel matrix by randomly sampling sub-blocks, with embedding coherence ensured by $QR$-decomposition.
	\item Grouping of reconstructed components using hierarchical clustering. In this case, the comparison criterion is the average distance of each component to the correlation matrix of all the components.
\end{itemize}

	To validate our approach, we compared it with the canonical-SSA method using real data sets that are well known in both geophysics and biophysics (\cf Section \ref{secIII}). Specifically, these datasets included the polar motion and thermal measurements in trees. For these two examples, we wanted to address two key points: first, a good correspondence between the patterns extracted by both methods, in terms of both phase and amplitude; and second, the ability to obtain comparable results within a reasonable time frame without downsampling the data.
	
	Regarding the first question, which focuses on the motion of the polar motion (\cf Section \ref{sec:polar}), the results obtained using canonical and pragmatic-SSA are more than comparable. The three components of the polar motion, the Chandler wobble, the annual oscillation and the Markowitz-Stoyko drift, were successfully identified in the main eigentriplets and reconstructed almost identically with both methods.
	
	The only notable discrepancy concerns the reconstruction of the Chandler component in Figure \ref{fig:04b}, where a significant difference appears from the 2000s onwards. This discrepancy is probably due to the fact that, as mentioned above, the Chandler period has slowed down during this period (from $\sim$1.17-1.19 years to the current 1.30 years).	We do not believe that this difference is due to edge effects for the following reasons,
\begin{enumerate}
	\item The algorithm was applied to an extended signal, so if edge effects were present, they would affect the duplicated segments that we remove at the end of the algorithm.
	\item For example, in Figure \ref{fig:05c}, there are no edge effects in the reconstruction of the temperature variations in the oak during the period February to May 2023 for either method.
\end{enumerate}	

	Regarding the second point, which concerns computational time and feasibility, we tested this in Section \ref{sec:trees} with our biological data. There are $\pi\times 10^{7}$ seconds in a year. The longest segment we analyzed, from February to mid-May 2023 (\cf Figures \ref{fig:05b} and \ref{fig:05c}), covers about $\dfrac{\pi}{4}\times 10^{7}$  seconds, resulting in a substantial Hankel matrix. Without downsampling, we obtained waveform patterns comparable to those in the previous section. In addition, we ensured the stability of the phase reconstruction even though we duplicated the edge segments of the signal (\cf Figure \ref{fig:05c}).
	
	We are quite happy with our pragmatic SSA approach, which compares well with canonical-SSA, although it is not perfect. There are still several points for improvement, the most important being the quality of the reconstruction for a given component. For example, in Figure \ref{fig:05b}, there are notable differences between the two approaches in the temporal evolution of the diurnal envelope from July to November 2023. It is difficult to favor one method over the other.
	
	In the case of the canonical-SSA, the raw signals had to be downsampled by a factor of 1000 (\ie one point every 0.2 hours, which is sufficient to evaluate the diurnal component) in order to obtain a signal size of the order of $10^4$ points. This downsampling makes the denationalization of the Hankel matrix feasible within a reasonable time on a standard computer. Because of this downsampling, in principle, the two methods do not analyze the same information. The matching was done on the basis of similar eigenvalues (Hilbert pairs). However, as noted in Section \ref{secII.1}, it is possible that parts of the diurnal information ended up in different eigentriplets, making the matching process inherently challenging.
	
	In our proposed approach, we do not downsample the signal. Instead, we iteratively approximate the SVD of the Hankel matrix block by block, working within the framework of truncated SVD. In addition, component matching is automated by hierarchical clustering, which in the most commonly used method is based on the distance between the reconstructed components and the correlation matrix of these components. However, this distance metric may not be the most appropriate for our biophysical data and is unlikely to be generalisable to all classes of signals. Further refinement of the matching process remains an important consideration for future work, For this purpose, the dendrogram we introduced (\cf Figure \ref{fig:04e}) will be of great utility.
	
\appendix
\section{An Example of Implementation in Matlab\label{app:A}}
Here is an example of a Matlab implementation of our pragmatic SSA. The functions in bold are Matlab's native functions, while the functions in red are our custom functions used to pedagogically segment the algorithm,

\begin{lstlisting}
function ssa = ssaDecompo(signal, L, rank, energyThreshold)
    extendedSignal      = [flipud(signal(1:L)); signal; flipud(signal(end-L+1:end))];
    ssa                 = ssaHankelRandomized(extendedSignal, L, rank, 4);
    singularValues      = diag(ssa.Lambda);
    cumulativeEnergy    = cumsum(singularValues.^2) / sum(singularValues.^2);
    numComponents       = find(cumulativeEnergy >= energyThreshold, 1);
    disp(['Total number of retained eigenvectors = ' num2str(numComponents)])
    ssa.group           = (1:numComponents)';
    ssa                 = ssaReconstruct(ssa);
    ssa.component       = ssa.component(L+1:end-L, :);
    ssa                 = ssaClustering(ssa);
end
\end{lstlisting}

	The first function is \textit{ssaDecompo()}. First, we replicate the edge segments of the signal with a size equal to the analysis window  using the \textit{flipud()} function. Next, we execute the block corresponding to the randomized SVD using the \textit{ssaHankelRandomized()} function, specifying the rank for the truncation. Here the value of 4 corresponds to the number of iterations. Obviously, for a rank equal to , the approach loses its usefulness. Once the denationalization is complete, we retrieve the eigenvalues and then identify those whose cumulative values exceed our threshold (energyThreshold). This step corresponds to the block using the Matlab functions \textit{diag(), cumsum()} and \textit{find()}. Using our \textit{ssaReconstruct()} function, we reconstruct the signal component corresponding to each isolated eigentriplets in the Hankelization phase. We then remove the parts of these components corresponding to the duplicated segments. Finally, we perform the clustering block using the \textit{ssaClustering()} function.
\begin{lstlisting}
function ssa = ssaHankelRandomized(s, L, rank, numIter)
		H                = hankel(s(1:L), s(L:end));
		[UK, Lambda, UL] = randomizedSVD(H, rank, numIter);
		ssa.T            = H;
		ssa.UK           = UK;
		ssa.Lambda       = Lambda;
		ssa.UL           = UL;
		ssa.k            = L;
		ssa.signal       = s;
end
\end{lstlisting}

	In this second function, \textit{ssaHankelRandomized()}, we first construct the Hankel matrix (\textbf{H}) from the windowed signal of size $L$ using Matlab's \textit{hankel()} function. The \textit{randomizedSVD()} function replaces the canonical SVD. We then store the necessary quantities in different fields of a structure (ssa) for later use.
\begin{lstlisting}
function [U, S, V] = randomizedSVD(H, rank, numIter)
		[m, n]			= size(H);
		rank			  = min([rank, m, n]);
		Q					  = randn(m, rank);
		for i = 1:numIter
			Q = H * (H' * Q); 
		end
		[Q, ~]		= qr(Q, 0); 
		B					= Q' * H;   
		[U_tilde, S, V]		= svd(B, 'econ');
		U					= Q * U_tilde; 
end
\end{lstlisting}

	In the \textit{randomizedSVD()} function, we first determine the rank of the Hankel matrix. We then reduce the dimension to the given rank. Next, we generate a matrix Q (with a rank lower than\textbf{ H}) filled with normally distributed random numbers and iterate the algorithm. Using the qr() function, we orthonormalize \textbf{Q} and project the Hankel matrix onto it. We then perform a standard SVD. Finally, we resize the eigenvectors to their appropriate dimensions in preparation for reconstruction.

\newpage
\bibliographystyle{fchicago}
\bibliography{ssa_biblio.bib}
\end{document}